\documentclass[a4paper,12pt]{article}
\pdfoutput=1
\usepackage{epsfig}
\usepackage{amsmath}
\usepackage{amssymb}
\usepackage{amsfonts}
\usepackage{bbm}
\usepackage{fleqn}
\usepackage[footnotesize]{caption}
\usepackage{graphicx}
\usepackage{mathrsfs}
\usepackage[center,footnotesize,hang]{subfigure}

\def\D{\mathrm{d}}

\def\O{\text{O}}

\def\be{\begin{equation}}
\def\ee{\end{equation}}
\def\beq{\begin{equation}}
\def\eeq{\end{equation}}
\def\bea{\begin{eqnarray}}
\def\eea{\end{eqnarray}}

\def\eps{\epsilon}

\def\<{\left\langle}
\def\>{\right\rangle}

\def\be{\begin{equation}}
\def\ee{\end{equation}}
\def\beq{\begin{equation}}
\def\eeq{\end{equation}}
\def\bea{\begin{eqnarray}}
\def\eea{\end{eqnarray}}

\def\a{\alpha}
\def\b{\beta}

\def\d{\delta}

\def\g{\gamma}

\def\k{\kappa}

\def\m{\mu}
\def\n{\nu}

\def\s{\sigma}
\def\t{\tau}

\def\D{\Delta}

\def\G{\Gamma}

\def\O{\Omega}

\def\ve{\varepsilon}

\def\mtt{\widetilde{m}_2}

\def\be{\begin{equation}}
\def\ee{\end{equation}}

\def\ra{\rightarrow}

\def\bea{\begin{eqnarray}}
\def\eea{\end{eqnarray}}

\newcommand{\newc}{\newcommand}
\newc{\ol}{\overline}
\newc{\wt}{\widetilde}
\newc{\bs}{\boldsymbol}
\newc{\ma}{\mathcal}
\newc{\vl}{\langle}
\newc{\vr}{\rangle}

\newc{\sg}{S}
\newc{\ug}{U}
\newc{\tg}{T}

\allowdisplaybreaks[2]
\begin{document}
\bibliographystyle{OurBibTeX}

\title{\hfill ~\\[-30mm]
                  \textbf{
                  Successful $N_2$ leptogenesis with flavour coupling effects in realistic unified models }        }
\date{\today}
\author{\\[-5mm]
Pasquale Di Bari\footnote{E-mail: {\tt P.Di-Bari@soton.ac.uk}} \ and 
Stephen F. King\footnote{E-mail: {\tt king@soton.ac.uk}} 
\\ \\
\emph{\small School of Physics and Astronomy, University of Southampton,}\\
 \emph{\small Southampton, SO17 1BJ, United Kingdom}\\[4mm]}

\maketitle

\begin{abstract}
\noindent
{In realistic unified models involving so-called $SO(10)$-inspired patterns of Dirac
and heavy right-handed (RH) neutrino masses, the lightest right-handed
neutrino $N_1$ is too light to yield successful
thermal leptogenesis, barring highly fine tuned solutions,
while the second heaviest right-handed neutrino $N_2$
is typically in the correct mass range.
We show that flavour coupling effects in the Boltzmann equations may be crucial to the 
success of such $N_2$ dominated leptogenesis, by helping to ensure that 
the flavour asymmetries produced at the $N_2$ scale survive
$N_1$ washout.
To illustrate these effects we focus on $N_2$ dominated leptogenesis in an existing model,
the A to Z of flavour with Pati-Salam,
where the neutrino Dirac mass matrix may be equal to an up-type quark mass matrix
and has a particular constrained structure. The numerical results, supported by analytical insight, show that in order to achieve successful $N_2$ leptogenesis, consistent with neutrino phenomenology,
requires a ``flavour swap scenario'' together with 
a less hierarchical pattern of RH neutrino masses than naively expected, at the expense of some mild 
fine-tuning. In the considered A to Z model neutrino masses are predicted to be normal ordered, 
with an atmospheric neutrino mixing angle 
well into the second octant and the Dirac phase $\d\simeq 20^{\circ}$, a set of predictions that will be 
tested in the next years in neutrino oscillation experiments.
Flavour coupling effects may be relevant for other $SO(10)$-inspired unified models
where $N_2$ leptogenesis is necessary.
} 
 \end{abstract}
\thispagestyle{empty}
\vfill
\newpage
\setcounter{page}{1}

\section{Introduction}

It would be certainly desirable to extend the SM, realising a unified picture able to
solve the flavour problem, explaining masses and mixing parameters of quarks and leptons, 
and at the same time to provide solution to the cosmological puzzles. In particular leptogenesis \cite{fy} is an attractive way
to explain the matter-antimatter asymmetry of the Universe, that in terms of the baryon-to-photon
number ratio is given by \cite{planck}
\be
\eta_B^{\rm CMB} = (6.1 \pm 0.1) \times 10^{-10} \,  .
\ee
Leptogenesis is a cosmological application of the see-saw mechanism, a minimal extension of the SM
able to explain neutrino masses and mixing \cite{seesaw}. 
Despite an intense activity on various aspects of leptogenesis, 
there are not many definite realistic unified models that have been shown 
to lead to successful leptogenesis while explaining fermion masses, mixing and CP violation,
although in the case of $SU(5)$ it is certainly possible to 
achieve successful $N_1$ leptogenesis (for a recent example see e.g. 
\cite{Bjorkeroth:2015tsa} which uses the sequential dominance results 
for $N_1$ leptogenesis discussed in \cite{Antusch:2006cw}).

In this paper we are interested in $N_2$ leptogenesis in so-called $SO(10)$-inspired models
with type I seesaw. 
For definiteness, we investigate the possibility that the A to Z Pati-Salam model proposed in \cite{A2Z}
(see also \cite{Bjorkeroth:2014vha})
can not only describe neutrino masses and mixing but also attain the correct value of the
matter-antimatter asymmetry with leptogenesis. The right-handed (RH) neutrino mass spectrum 
in this model is very hierarchical, typical of $SO(10)$-inspired models
(assuming type I seesaw). In this way the lightest RH neutrino $N_1$ is too light to generate a sizeable
asymmetry \cite{di} while on the other hand the next-to-lightest RH neutrino is heavy enough to be able potentially to generate the correct
asymmetry realising the so called $N_2$ dominated scenario \cite{geometry,N2dominated,riotto,decription} or simply $N_2$ leptogenesis.  
However in general it is non-trivial to be able to find a set of values of the parameters of the model
for the lightest RH neutrino wash-out to be negligible, in such a way that the asymmetry
generated by the next-to-lightest neutrinos survives, while at the same time  producing values of the 
neutrino parameters compatible with the experimental results.  

The non-trivial requirements of $N_2$ leptogenesis 
may be somewhat relaxed by taking into account additional flavour coupling effects 
in the Boltzmann equations, or more generally the kinetic mixing terms, 
that may transmit part of the initially produced flavour asymmetry from one particular flavour to other flavours. This increases the chances that the asymmetry generated in a particular flavour at the $N_2$ scale
may survive the $N_1$ washout.
The additional flavour coupling effects, which are not usually considered in the literature, arise mainly from the 
fact that lepton asymmetry produced at the $N_2$ scale into left-handed lepton doublets 
is also accompanied  by hyper-charge asymmetry into Higgs 
asymmetry, giving the dominant effect, baryons asymmetries into quarks and lepton asymmetries into right-handed charged leptons, 
which are then transmitted to the other flavours 
which couple to these particles \cite{flcoupling}.  This results in new contributions to the total asymmetry that 
in a traditional $N_1$-dominated scenario would give just some small corrections (${\cal O}(10\%)$) \cite{abada}. 
However, in a $N_2$-dominated scenario they can become dominant if, contrarily to the usual dominant terms produced in a flavour
that is strongly washed by the lightest RH neutrinos, they are produced into a flavour that escapes the lightest RH neutrino wash-out.
In this way  the so called ``flavour swap scenario'' is realised \cite{Antusch:2010ms}. 

In the A to Z of Flavour with Pati-Salam \cite{A2Z}, 
all of these features are exemplified.
In particular, the $N_1$ scale is too light
to generate asymmetries thermally, while the flavour asymmetry produced at the $N_2$ scale, namely the $\tau$
flavour, is effectively washed out at the $N_1$ scale for all ranges of parameters consistent with the experimentally
acceptable neutrino masses and mixings. 
In this example, flavour coupling effects come to the rescue,
effectively transmitting part of the $\tau$ asymmetry into (a linear combination of) electron and muon asymmetries
at the $N_2$ scale, completely (electron) or partly (muon) surviving washout at the $N_1$ scale.
These features necessarily arise from the rigid structure of the Yukawa and Majorana matrices enforced by the model,
leading to fairly precise predictions for PMNS parameters which can be compared to experiment.
It has been already shown that the model can simultaneously fit both lepton and quark parameters. 
Here for simplicity we will focus on the leptonic sector and consider non-supersymmetric leptogenesis. 

The layout of the remainder of the paper is as follows. In Section 2 we review neutrino masses in the A to Z model.
In Section 3 we fit  the neutrino parameters without imposing successful leptogenesis. In Section 4 we show how within
a traditional calculation of the asymmetry ignoring flavour coupling effects one cannot find any good fit both to neutrino
parameters and to the measured value of the asymmetry simultaneously. In Section 5 we show how flavour coupling rescues the model.
In Section 6 we provide an analytical insight on the found solution within more general context of $SO(10)$-inspired
leptogenesis highlighting different aspects both of the specific solution within the A to Z model and more generally 
of $SO(10)$-inspired models. In Section 7 we draw our conclusions. 

\section{Neutrino masses in the A to Z model}

The lowest order lepton Yukawa matrices
(in LR convention) and heavy Majorana mass matrix $M_R$
are predicted by the model just below the high energy Pati-Salam breaking scale $\sim $ few $\times 10^{16}$ GeV. 
The charged lepton Yukawa matrix of the model is diagonal to excellent approximation, and 
the neutrino Lagrangian in this basis is given by,
\beq
-{\cal L}= \overline{N}_LY^{'\nu}N'_R+N^{'T}_RM'_RN'_R + H.c.
\eeq
with the neutrino Yukawa and Majorana matrices \cite{A2Z}, 
\beq
Y^{'\nu}  = \begin{pmatrix}  
0 & be^{-i3\pi/5}   & 0 \\ 
ae^{-i3\pi/5} & 4be^{-i3\pi/5} & 0\\  
ae^{-i3\pi/5}  & 2be^{-i3\pi/5} & ce^{i\phi}\end{pmatrix},
\ \ 
 M'_R  =   \begin{pmatrix} M'_{11}\, e^{2i \xi} & 0 & M'_{13}\, e^{i \xi} \\ 0 &  
M'_{22}\, e^{i \xi } &0 \\  M'_{13}\,e^{i \xi } & 0& M'_{33} \end{pmatrix}  \,   ,
\label{Ynu}
\eeq
where $M'_{11}, M'_{13}, M'_{22}, M'_{33}$ are positive and real.
Note that for simplicity it was also assumed in \cite{A2Z} that $M'_{13}=0$ and $\phi=0$,
while here we shall allow these parameters (generally present in the model) to take non-zero values.
Generally the model predicts 
$M'_{11}\ll M'_{22} \ll M'_{33}$ and $M'_{13}\sim  M'_{22} $.
Without any tuning of parameters, the model predicts the typical values,
\beq\label{typical}
M'_{11} \sim 10^5 \ {\rm GeV}, \ M'_{22} \sim M'_{13} \sim 10^{10}\ {\rm GeV}, \  M'_{33}\sim 10^{15}\  {\rm GeV},
\eeq
while 
\beq\label{typical2}
a:b:c \sim m_{\rm up}:m_{\rm charm}:m_{\rm top} \sim 10^{-6}: 10^{-3} : 1   \,   .
\eeq
However, successful leptogenesis requires some fine-tuning, leading to a more compressed
spectrum of right-handed neutrino masses, as we shall see.
Also $\xi$ is chosen to be one of the complex fifth roots of unity: $\xi=0, \pm 2\pi/5 , \pm 4\pi/5 $.
We shall consider all cases.

We perform a unitary transformation $U_R$ to the flavour basis where the right-handed Majorana mass matrix
is diagonal with real, positive eigenvalues, 
\bea
{\cal L}&=& \overline{N}_LY^{'\nu}\,U_R\, U^{\dagger}_R \, N'_R
+N^{'T}_R \, U^*_R \, U^{T}_R \, M'_R \, U_R \, U^{\dagger}_R \, N'_R +H.c.\\
&=& \overline{N}_L \, Y^{\nu}\, N_R
+N^{T}_R\, M_R \, N_R +H.c.
\eea
where 
\be
N_R=U^{\dagger}_R \, N'_R, \ \  Y^{\nu}=Y^{'\nu}\, U_R, \ \ M_R=U^{T}_R \, M'_R \, U_R= {\rm diag}(M_1, M_2, M_3).
\eeq
We parametrise the unitary matrix, assuming small angle rotations, approximately as
\beq
U_R = 
\begin{pmatrix}
1& 0 & R_{13}e^{i\phi_{13}}\\
0 & 1 & 0 \\
-R_{13}e^{-i\phi_{13}} & 0 & 1
\end{pmatrix}
\begin{pmatrix}
e^{i\phi_{11}} & 0 & 0 \\
0 & e^{i\phi_{22}} & 0 \\
0 & 0 & e^{i\phi_{33}}
\end{pmatrix}  \,   ,
\eeq
where the first matrix factor diagonalises the right-handed neutrino mass matrix and the second factor ensures that 
the right-handed neutrino masses $M_i$ are real and positive.
Thus the form of the neutrino Yukawa matrix in the flavour basis is
$Y^{\nu}=Y^{'\nu}\, U_R$,
\beq
Y^{\nu} \approx 
\begin{pmatrix}  
0 & be^{-i3\pi/5}e^{i\phi_{22}}   & 0 \\ 
ae^{-i3\pi/5}e^{i\phi_{11}} & 4be^{-i3\pi/5}e^{i\phi_{22}} & a\,R_{13}\,e^{i( \phi_{33}-\xi-3\pi/5)} \\  
ae^{-i3\pi/5}e^{i\phi_{11}}-cR_{13}e^{i\phi}e^{i(\phi_{11}-\phi_{13})} & 2be^{-i3\pi/5}e^{i\phi_{22}} & ce^{i\phi}e^{i\phi_{33}}  
\end{pmatrix} \,   .
\eeq
The parameters of $U_R$ are determined from the requirement that 
\beq
M_R=U^{T}_RM'_RU_R= D_M \equiv {\rm diag}(M_1, M_2, M_3).
\label{MR}
\eeq
In particular, the matrix elements of the diagonal $M_R$ must satisfy:
\bea
(M_R)_{11}&=&M_1\approx e^{2i(\phi_{11}-\phi_{13}) } (e^{2i(\xi+\phi_{13})} M'_{11} - 2 e^{i(\xi+\phi_{13})} M'_{13}R_{13} 
+M'_{33}R^2_{13}     ) \,  ,\nonumber \\
(M_R)_{13}&=&0\approx e^{i(\phi_{11}-\phi_{13}+\phi_{33}) } (e^{2i(\xi+\phi_{13})} M'_{11}R_{13} - 
 e^{i(\xi+\phi_{13})} M'_{13}(R^2_{13}-1) -M'_{33}R_{13}     )  \,   ,\nonumber \\
 (M_R)_{22}&=&M_2\approx e^{2i\phi_{22} }e^{i\xi}M'_{22} \,   ,\nonumber \\
 (M_R)_{33}&=&M_3\approx e^{2i\phi_{33}}  (M'_{33}+e^{2i(\xi+\phi_{13})} M'_{11}R^2_{13}  + 2 e^{i(\xi+\phi_{13})} M'_{13}R_{13} 
  )  \,   .\nonumber \\
\eea
From these conditions (remembering that $M_i$ are real and positive) we find:
\beq
R_{13}\approx \frac{M'_{13}}{M'_{33}}, \ \phi_{13}= -\xi , \ \phi_{11}=
-\xi+{1\over 2}\,{\rm Arg}\left[M'_{11}-{M_{13}^{'2} \over M'_{33}}\right] \  , \  \phi_{22}\approx -\xi/2 , \ 
\phi_{33}\approx 0 \, ,  
\eeq
leading to:
\bea
M_1&\approx&   \left| M'_{11} - 2  M'_{13}R_{13}   +M'_{33}R^2_{13}\right| \approx    \left| M'_{11}- {M_{13}^{'2} \over M'_{33}} \right| \,   , \nonumber \\
M_2&\approx& M'_{22} \nonumber   \,   , \\
M_3&\approx& M'_{33}+ M'_{11}R^2_{13}  + 2 M'_{13}R_{13} \,   . \nonumber \\
\eea

Thus the neutrino Yukawa matrix in the flavour basis is given by,
\beq
\label{Ynufl}
Y^{\nu} \approx 
\begin{pmatrix}  
0 & be^{-i\,(\xi/2+3\pi/5)}  & 0 \\ 
a\,e^{-i(\xi+3\pi/5)} & 4be^{-i\,(\xi/2+3\pi/5)} & a\,R_{13}\,e^{-i( \xi+3\pi/5)} \\  
a\,e^{-i\,(\xi+3\pi/5)}(1-\gamma) & 2be^{-i(\xi/2+3\pi/5)} & c\,e^{i\phi}
\end{pmatrix} \,  ,
\eeq  
where 
\beq\label{gamma}
\gamma \approx (c/a)R_{13}e^{i(\phi +\xi+3\pi/5)} \,   .
\eeq
The Dirac neutrino masses, the eigenvalues of the neutrino Dirac mass matrix $m_{\nu}^D = v\,Y^{\nu}_{LR}$, are given by
\beq
m^D_{\nu1} \sim a \, v / \sqrt{17}, \ \ m^D_{\nu2}\sim \sqrt{17} \, b\, v  \ \ 
m^D_{\nu3} \sim c \, v  \,  ,
\label{Dirac}
\eeq
where the approximations are quite precise given the working assumptions
leading to the typical values in Eq.~(\ref{typical2}).

For the up-type quark matrix we have two options,
\beq\label{CASEA}
Y^{u}  = \begin{pmatrix}  
0 & be^{-i3\pi/5}   & 0 \\ 
ae^{-i3\pi/5} & 4be^{-i3\pi/5} & 0\\  
ae^{-i3\pi/5}  & 2be^{-i3\pi/5} & c\, e^{i\phi}\end{pmatrix}
\hspace{20mm} \mbox{CASE A}
\eeq
and
\beq\label{CASEB}
 Y^{u}  = \begin{pmatrix}  
0 & \frac{1}{3}\, be^{-i3\pi/5}   & 0 \\ 
ae^{-i3\pi/5} & \frac{4}{3}\,be^{-i3\pi/5} & 0\\  
ae^{-i3\pi/5}  & \frac{2}{3}\,be^{-i3\pi/5} & 3\, ce^{i\phi}\end{pmatrix}  
\hspace{17mm} \mbox{CASE B} \, .
\eeq
These are simply related to the neutrino Yukawa matrix in the original basis,
$Y^{' \nu}$, by Clebsch relations for CASE B, while we have simply
$Y^{u} =Y^{'\nu} $ for CASE A (in the basis of Eq.(\ref{Ynu}))
which is just the minimal
$SO(10)$-like expectation that the Dirac neutrino mass matrix is identically equal to the
up-type quark mass matrix. Notice that in the A to Z model there are ``texture zeroes'' in the
(1,1), (1,3) and (2,3) entries of the Yukawa matrices above that will play a role
in leptogenesis considerations.

The Dirac neutrino masses are simply related to the up-type quark masses, 
depending on the choice of model,
\beq\label{CASEAbis}
m^D_{\nu1} = m_{\rm up}, \ \ m^D_{\nu2} = \, m_{\rm charm}, \ \ 
m^D_{\nu3} =  m_{\rm top} \hspace{15mm} \mbox{\rm CASE A}
\label{upA}
\eeq
and
\beq\label{CASEBbis}
m^D_{\nu1}\approx m_{\rm up}, \ \ m^D_{\nu2}\approx  3\, m_{\rm charm}, \ \ 
m^D_{\nu3}\approx  {1\over 3} \, m_{\rm top} \hspace{10mm} \mbox{\rm CASE B}    \,  .
\label{upB}
\eeq
In this way the  (real) parameters $a$, $b$ and $c$ are simply 
determined from the values of the up quark masses at the grand-unified scale
\footnote{We use the values $m_{\rm top} = 100\,{\rm GeV}$, $m_{\rm charm}=400\,{\rm MeV}$ and 
$m_{\rm up} = 1\,{\rm MeV}$. \cite{quarks}}
so that the (real) input parameters are: $M'_{11},M'_{13},M'_{22},M'_{33}$ and phase $\phi$.
The phase $\xi$ is restricted to be one of the complex fifth roots of unity: $\xi=0, \pm 2\pi/5 , \pm 4\pi/5 $.

\subsubsection*{Seesaw mechanism in the flavour basis}

Using the see-saw formula 
\be\label{seesaw}
m^{\nu}=-v^2 \, Y^{\nu} \,  M_R^{-1} \,  Y^{\nu T}
\ee 
in the flavour basis we find the neutrino mass matrix $m^{\nu}$, 
\beq
m^{\nu} 
\approx m_a\, e^{i\phi_a} \begin{pmatrix} 0 & 0 & 0 \\ 0 & 1 & (1-\gamma) \\ 0 & (1-\gamma) & (1-\gamma)^2 \end{pmatrix} 
+ m_b\, e^{i\phi_b}\begin{pmatrix} 1 & 4 & 2 \\ 4 & 16 & 8 \\ 2 & 8 & 4  \end{pmatrix}
+m_c \, e^{i\phi_c} \begin{pmatrix} 0 & 0 & 0 \\ 0 & 0 & 0 \\ 0 & 0 & 1 \end{pmatrix} \,   ,
\label{seesaw3}
\eeq
\beq
m_a=\frac{a^2v^2}{ {M}_1}, \ m_b=\frac{b^2v^2}{ {M}_2}, \ m_c=\frac{c^2v^2}{ {M}_3}  \,   ,
\label{abc}
\eeq

\beq
\phi_a=-2(\xi+3\pi/5), \ \phi_b=-2(\xi/2+3\pi/5), \ \phi_c=2\phi   \,   .
\label{phiabc}
\eeq
From Eqs.(\ref{Dirac}),(\ref{abc}) and (\ref{upA})  or (\ref{upB})  we find,
\beq\label{massrelations}
m_a\sim 17\frac{m_{\rm up}^2}{M_1}, \ m_b \sim  (9)\frac{m_{\rm charm}^2}{17M_2}, 
\ m_c\sim \frac{m_{\rm top}^2}{(9)M_3}  \,   ,
\eeq
where the factors in brackets apply to CASE B, and these factors are simply unity for CASE A.
Thus the three right-handed neutrino masses $M_1$, $M_2$, $M_3$ may be determined
from $m_a$, $m_b$, $m_c$.


The Majorana neutrino mass
matrix $m^\nu$, defined by 
$\mathcal{L}_\nu=-\tfrac{1}{2} m^\nu \overline \nu_{\mathrm{L}} 
\nu^{c}_{\mathrm{L}}$ + h.c., is diagonalised by
\begin{eqnarray}\label{eq:DiagMnu}
U_{\nu_\mathrm{L}} \,m^\nu\,U^T_{\nu_\mathrm{L}} =
\left(\begin{array}{ccc}
\!m_1&0&0\!\\
\!0&m_2&0\!\\
\!0&0&m_3\!
\end{array}
\right)\! .
\end{eqnarray}  
The PMNS matrix is then given by
\begin{eqnarray}
U_{\mathrm{PMNS}} = U_{e_\mathrm{L}} U^\dagger_{\nu_\mathrm{L}}\; .
\end{eqnarray}
We use a standard parameterization 
$
U_{\mathrm{PMNS}} = R_{23} U_{13} R_{12} P
$ 
in terms of $s_{ij}=\sin (\theta_{ij})$,
$c_{ij}=\cos(\theta_{ij})$, the Dirac CP violating phase $\delta^l$ and
further Majorana phases contained in $P={\rm diag}(e^{i\frac{\beta_1}{2}},e^{i\frac{\beta_2}{2}},1)$.
The standard PDG parameterization  \cite{PDG} differs slightly due to the definition of Majorana phases which are by given by $P_{\rm PDG}={\rm diag}(1,e^{i\frac{\alpha_{21}}{2}},e^{i\frac{\alpha_{31}}{2}})$.
Evidently the PDG Majorana 
phases are related to those in our convention by $\alpha_{21}=\beta_2^l-\beta_1^l$ and $\alpha_{31}=-\beta_1^l$,
after an overall unphysical phase is absorbed by $U_{e_\mathrm{L}}$.

For example, using the input parameters $\xi=4\pi/5$,
$\phi=\gamma=0$ and 
\beq
m_a = 0.034 \ {\rm eV}, \
m_b = 0.002  \ {\rm eV},
m_c=0.002 \ {\rm eV}, 
\eeq
we find for the CASE A the following values for the physical neutrino masses,
\beq
m_1 = 0.00034 \ {\rm eV}, \
m_2=0.0086 \ {\rm eV}, \
m_3=0.050 \ {\rm eV}. 
\eeq
and the lepton mixing parameters,
\beq
\theta_{12}=34.0^\circ, \ 
\theta_{13}=9.1^\circ, \ 
\theta_{23}=39.7^\circ, \ 
\delta=260^\circ, \ 
\beta_1=321^\circ, \
\beta_2=75^\circ \,  .
\eeq

\section{Fitting neutrino parameters}

In this Section we perform a quantitative numerical analysis.
We randomly generate
a set of values of the five (continuous) parameters of the model ($\underline{y} \equiv M'_{11}, M'_{22}, M'_{33},M'_{13}, \phi$)
for each of the five choices of the discrete parameter $\xi$ and from these we calculate the nine
low energy neutrino parameters in $m_{\nu}$.
A recent  global analysis \cite{nufit} finds the values shown below (with $1\s$ errors).

The solar mass squared difference: 
\be
\Delta m^{2}_{12} \equiv m^2_2 -m^2_1 = 7.50^{+0.19}_{-0.17}\times 10^{-5}\,{\rm eV}^2  .
\ee
The atmospheric mass squared difference, 
respectively for normal ordering (NO) and for inverted ordering (IO):
\bea
\Delta m^2_{31} & \equiv &  m^2_3 -m^2_1 = 2.457^{+0.047}_{-0.047}\times 10^{-3}\,{\rm eV}^2 \hspace{3mm} \mbox{\rm and} \\ \nonumber
\hspace{3mm} \Delta m^2_{32} & \equiv & m^2_3 -m^2_2 =  -2.449^{+0.048}_{-0.047}\times 10^{-3}\,{\rm eV}^2  \,  .
\eea
The solar mixing angle:
\be
\theta_{12} = \left.33.48^{\circ}\right.^{+0.78^{\circ}}_{-0.75^{\circ}}  \,   .
\ee
The reactor mixing angle:
\be
\theta_{13} =  \left. 8.50^{\circ}\right.^{+0.20^{\circ}}_{-0.21^{\circ}} \hspace{6mm} \,  .
\ee
The atmospheric mixing angle, respectively for NO and for IO:
\be\label{atmangle}
\theta_{23} = \left. 42.3^{\circ}\right.^{+3.0^{\circ}}_{-1.6^{\circ}} \hspace{6mm} \mbox{\rm and} 
\hspace{4mm} \theta_{23} = \left.49.5^{\circ}\right.^{+1.5^{\circ}}_{-2.2^{\circ}} \,  .
\ee
The Dirac phase, respectively for NO and for IO: 
\be\label{diracphase}
\delta = \left. 306^{\circ}\right.^{+39^{\circ}}_{-70^{\circ}} \hspace{6mm} \mbox{\rm and} 
\hspace{4mm} \delta = \left.254^{\circ}\right.^{+63^{\circ}}_{-62^{\circ}} \,  .
\ee
We see that, for the atmospheric mixing angle, within $1\sigma$, 
there are currently two solutions, one in the first octant for NO and one in the second octant for IO,
though the alternative choice of octant is in both cases only disfavoured at only $\sim 1.4\,\s$.
Note also that, within $3\s$, the Dirac phase $\delta$ can have any value. 

The four parameters $\Delta m^2_{12}, \Delta m^2_{23}, \theta_{12}$ and $\theta_{13}$, are measured 
quite accurately and precisely and their distributions are very well approximated by Gaussian distributions. 
On the other hand the atmospheric mixing angle is not only much less precisely measured and not 
Gaussianly distributed,  but is also 
affected by larger systematic uncertainties and in particular different global analyses 
exclude  maximal mixing with different statistical significance, in any case below $2\s$,
so that its determination, and in particular the deviation from maximality, should be still regarded  as
quite unstable \cite{nufit,global}.   

We have determined our best fit values minimising the quantity
\be\label{chi2}
\chi^2 \equiv  \sum_{ i=1}^{N}  \, p^2_i  \,  , \hspace{10mm}  p_i \equiv {X_i^{\rm th}(\underline{y})-\bar{X}_i \over \s_{X_i}}  \,  ,
\ee
where the $X_i = \bar{X}_i \pm \s_{X_i}$'s are the $N=5$   experimental parameters that we fit
(in next Sections we will include the matter-antimatter asymmetry so that $N$ rises to $6$) and
$X_i^{\rm th}(\underline{y})$ are their predicted values depending on the  theoretical parameters
of the model $\underline{y}=(M'_{11}, M'_{13}, M'_{22}, M'_{33}, \phi; \xi, \mbox{\rm J})$ 
($\xi =0,\pm 4\pi/5,\pm 2\pi/5$, J=CASE A, CASE B are discrete parameters).
Since the Dirac phase $\delta$ can have any value within $3\s$, and has certainly not
a Gaussian distribution, we decided not to include it in the fit not to risk to prematurely
exclude potential solutions but we will comment on how this would change including $\d$.

For the same reason for the atmospheric mixing angle we  use very conservatively 
$\theta_{23}=45.9^{\circ} \pm 3.5^{\circ}$ both for NO and IO (we have basically
taken as a central value the average between the two minima and as error 
their halved separation). 
We could also have  performed the fits distinguishing two different 
ranges for $\theta_{23}$, one in the lower octant for NO and one in the higher octant for IO, 
but these are still too weak hints. Our choice  cuts too low or too high values for $\theta_{23}$, 
as established by all experiments even singularly taken, but
it still treats the  maximal mixing  value as perfectly allowed and does not favour any of the two octants 
on the other, since, as already discussed, the hint coming from  current experimental data is still too weak and unstable. 
In this way  any emerging possible preference either for  lower or for higher octant, as for 
 maximal mixing, can be uniquely ascribed to the  model itself and is not hidden by still unstable measurements. 
 
Within the approximation that $\theta_{23}$ distribution is also Gaussianly distributed,
the defined quantity  $\chi^2$ has  a truly $\chi^2$-distribution,
being the sum of squares of  independent Gaussian variables 
and the $p_i$'s are the associated pulls. 

 Notice that the number of (continuous) theoretical parameters is equal to the number of experimental parameters (five)
and, therefore the number of degrees of freedom vanishes. This implies that the minimisation of $\chi^2$
can only provide best fit values (and prediction on the low energy phases) but cannot be 
regarded as a goodness of fitness of the model, since even vanishing $\chi^2$ values  could be potentially obtained
with a fine tuned choice of the theoretical parameters $\underline{y}$ independently of the 
values of the experimental values. However, a preliminary indication of the goodness of the fitness is given by a 
comparison of the predicted value of the Dirac phase with the current best fit value Eq.~(\ref{diracphase}), 
though, as we said, we certainly need  a more precise measurement of $\d$ to draw firmer conclusions. 

In Table 1 and Table 2 we summarise the results of our analysis. 
We indicate the Majorana phases both  in the convention 
${\rm diag}(e^{i\,{\beta_1\over2}}, e^{i\,{\beta_2\over2}},1)$
and in the (PDG) convention ${\rm diag}(1,e^{i\,{\alpha_{21}\over2}},e^{i\,{\alpha_{31}\over2}})$.
The table refers to NO, since we found 
that the model cannot reproduce the neutrino mixing parameters for IO. In particular it badly fails in reproducing the neutrino mass spectrum. 

Table 1 refers to  CASE A (cf. eq.~(\ref{CASEA})) while Table 2  to CASE B
(cf. eq.~(\ref{CASEB})).  For both cases we show the results for $\xi = \pm 4\pi/5$ and for $\xi=0$ since for $\xi=\pm 2\pi/5$
we could not find any fit with a value of $\chi^2 < 100$ so that they can be
basically considered ruled out. 
\begin{table}
	\centering
\begin{tabular}{|c| c |c| c|  }
				      \hline
CASE  & \multicolumn{3}{c|}{A}  \\
\hline \hline
$ \xi $  &  
\multicolumn{1}{c|}{+$4\pi/5$} &  
\multicolumn{1}{c|}{$0$} & 
\multicolumn{1}{c|}{-$4\pi/5$}    \\
\hline \hline
$\chi^2_{\rm min}$ & 4.40   &  22.8  & 3.63    	\\
                    \hline
$m_1/{\rm meV}$    & 0.20    & 0.021  & 0.022          \\
			\hline                                        
$m_2/{\rm meV} \; (p_{\D m^2_{12}})$       & 8.66 (+0.002)    &  8.62 (-0.38)   &  8.69 (+0.25)       \\
			\hline 
$m_3/{\rm meV}  \; (p_{\D m^2_{13}})$    & 49.9 (+0.76)    & 48.9 (-1.34)  &  49.8  (+0.47)       \\
			\hline   
$\theta_{12}/^{\circ} \; (p_{\theta_{12}})$     &  33.5 (+0.04)   & 35.3 (2.37)  & 33.0  (-0.66)       \\
			\hline   
$\theta_{13}/^{\circ} \; (p_{\theta_{13}})$     &  8.37 (-0.64)   & 8.08  (-2.05) & 8.42 (-0.37)    \\
			\hline   
$\theta_{23}/^{\circ} \; (p_{\theta_{23}})$    &  39.2 (-1.85)   &  34.3 (-3.3)  & 39.9 (-1.66)     \\
			\hline   
$\d/^{\circ}$                     &  251   &  180 &  109      \\
			\hline
$\b_1/^{\circ}$                     &  225   & 338  &  173       \\
			\hline
$\b_2/^{\circ}$                     &  82   & 175  &  279 \\
			\hline   			 	      
$\a_{21}/^{\circ}$                     &  217   &  197 & 106      \\
			\hline   
$\a_{31}/^{\circ}$                     &  135   &  22 &  187      \\
			\hline   \hline	
$M'_{11}/{\rm GeV}$                     &  $6.5\times 10^5$     &  $6.6\times 10^5$ &  $2.2 \times 10^6$      \\
			\hline 			
$M'_{22}/{\rm GeV}$                     & $5.0\times 10^9$    &  $5.1\times 10^9$  & $5.0\times 10^9$       \\
			\hline	
$M'_{33}/{\rm GeV}$                     & $7.2\times 10^{15}$     & $8.6\times 10^{16}$  &  $6.8\times 10^{16}$    \\
			\hline	
$M'_{13}/{\rm GeV}$                     & $3.2 \times 10^{10}$    &  $3.0 \times 10^{11}$ &  $3.4 \times 10^{11}$       \\
			\hline
$M'_{13}/M'_{22}$                         & 6.4 & 59.7  & 69       \\
			\hline					
			
$\phi/\pi$                     & 1.27    &  1.80 &  1.64      \\
			\hline		
$M_{1}/{\rm GeV}$                     & $5.1\times 10^5$    &  $4.2\times 10^5$ & $4.9\times 10^5$       \\
			\hline	
$M_{2}/{\rm GeV}$                     &$5.0\times 10^9$&  $5.1\times 10^9$ &  $5.0\times 10^9$     \\
			\hline	
$M_{3}/{\rm GeV}$                     & $7.2\times 10^{15}$    &  $8.6 \times 10^{16}$  & $6.8 \times 10^{16}$      \\
			\hline	
			\hline
                    \end{tabular} 
			\caption{Results for the case A (NO and no leptogenesis).}
		\label{table}
\end{table}

\begin{table}
	\centering
\begin{tabular}{|c| c |c| c|  }
				      \hline
CASE  & \multicolumn{3}{c|}{B}  \\
\hline \hline
$ \xi $  &  
\multicolumn{1}{c|}{+$4\pi/5$} &  
\multicolumn{1}{c|}{$0$} & 
\multicolumn{1}{c|}{-$4\pi/5$}    \\
\hline \hline
$\chi^2_{\rm min}$ &  7.84  &  9.32  & 5.81    	\\
                    \hline
$m_1/{\rm meV}$    &  0.008   & 3.0  & 0.004          \\
			\hline                                        
$m_2/{\rm meV} \; (p_{\D m^2_{12}})$       &  8.58 (-0.81)    &  9.25 (+0.89)   &  8.72 (+0.62)       \\
			\hline 
$m_3/{\rm meV}  \; (p_{\D m^2_{13}})$    & 49.6 (+0.05)    & 48.8 (-1.85)  &  50.3  (+1.53)       \\
			\hline   
$\theta_{12}/^{\circ} \; (p_{\theta_{12}})$     &  33.2 (-0.31)   & 32.3 (-1.49)  & 33.0  (-0.57)       \\
			\hline   
$\theta_{13}/^{\circ} \; (p_{\theta_{13}})$     &  8.93 (+2.10)   & 8.84  (+1.66) & 8.51 (+0.06)    \\
			\hline   
$\theta_{23}/^{\circ} \; (p_{\theta_{23}})$    & 40.0  (-1.63)   &  44.55 (-0.33)  & 39.9 (-1.66)     \\
			\hline   
$\d/^{\circ}$                     &  253  &  99 &  108     \\
			\hline
$\b_1/^{\circ}$                     &  335   & 66  &  170       \\
			\hline
$\b_2/^{\circ}$                     &   84  & 325  &  278 \\
			\hline   			 	      
$\a_{21}/^{\circ}$                     & 108    &  259 & 109      \\
			\hline   
$\a_{31}/^{\circ}$                     &  25  &  294 &  190      \\
			\hline   \hline	
$M'_{11}/{\rm GeV}$                     &  $7.2\times 10^5$     &  $2.9\times 10^5$ &  $6.6 \times 10^6$      \\
			\hline 			
$M'_{22}/{\rm GeV}$                     & $4.3\times 10^{10}$    &  $4.2\times 10^{10}$  & $4.4\times 10^{10}$       \\
			\hline	
$M'_{33}/{\rm GeV}$                     & $2.2\times 10^{16}$     & $9.7\times 10^{13}$  &  $4.3\times 10^{16}$    \\
			\hline	
$M'_{13}/{\rm GeV}$                     & $6.8 \times 10^{10}$    &  $7.5 \times 10^{9}$ &  $5.1 \times 10^{11}$       \\
			\hline	
$\phi/\pi$                     & 1.62    &  1.23 &  1.62      \\
			\hline		
$M_{1}/{\rm GeV}$                     & $5.1\times 10^5$    &  $2.9\times 10^5$ & $4.9\times 10^5$       \\
			\hline	
$M_{2}/{\rm GeV}$                     &$4.3\times 10^{10}$&  $4.2\times 10^{10}$ &  $4.4\times 10^{10}$     \\
			\hline	
$M_{3}/{\rm GeV}$                     & $2.2\times 10^{16}$    &  $9.7 \times 10^{13}$  & $4.3 \times 10^{16}$      \\
			\hline	
			\hline
                    \end{tabular} 
			\caption{Results for the CASE B (NO and no leptogenesis).}
		\label{table}
\end{table}

Let us  list the main results postponing some comments to the conclusions.
\begin{itemize}
\item We do not find any solution for IO, the model is unable to reproduce IO neutrino masses.
\item For $\xi =\pm 2\pi/5$ there are no solutions with $\chi^2_{\rm min}< 100$. 
\item Except for the case $\xi=0$, the best fit values are obtained for the CASE A.  
\item The best fit solutions that we obtained for NO, except for the case B with $\xi=0$, seem to point to
$\theta_{23}\sim 40^{\circ}$, certainly in the first octant and even below the current best fit value eq.~(\ref{atmangle})
hinted by current global analyses.
\item Taking into account the RH neutrino mixing parameter $M'_{13}$ improves all fits and one finds 
values of $\chi^2_{\rm min}(M'_{13}\neq 0) \simeq \chi^2_{\rm min}(M'_{13} = 0) -2$.  
\item Values of the Dirac phase close to the best fit value from global analyses (cf. eq.~(\ref{diracphase})) are attained for 
$\xi = + 4\,\pi/5$ (both in  CASE A and in CASE B). If further experimental data should further 
support these values ($\sim -90^{\circ}$), this would be the only surviving option. 
\end{itemize}

\section{Leptogenesis without flavour coupling} 

Let us now consider the calculation of the matter-antimatter asymmetry with leptogenesis.
The strongly hierarchical RH neutrino mass spectrum in the A to Z model discussed in the previous section,
with $M_1 \ll 10^9\,{\rm GeV}$ and $10^{12}\,{\rm GeV}\gg M_2 \gg 10^9\,{\rm GeV}$, necessarily points to a
$N_2$-dominated leptogenesis scenario \cite{geometry}  since  both the lighest and
the heaviest RH neutrino decays  produce a negligible asymmetry compared to the observed one.  In this case the $B-L$ asymmetry, in a 
portion of co-moving volume containing one RH neutrino in ultra-relativistic thermal equilibrium, can be calculated as
\cite{geometry,N2dominated}
\bea\label{twofl} \nonumber
N_{B-L}^{\rm lep, f}& \simeq &
\left\{\left[{K_{2e}\over K_{2\tau_2^{\bot}}}\,\ve_{2 \tau_2^{\bot}}\kappa(K_{2 \tau_2^{\bot}}) 
+ \left(\ve_{2e} - {K_{2e}\over K_{2\tau_2^{\bot}}}\, \ve_{2 \tau_2^{\bot}} \right)\,\kappa(K_{2 \tau_2^{\bot}}/2)\right]\,
\, e^{-{3\pi\over 8}\,K_{1 e}}+ \right. \\ \nonumber
& + &\left[{K_{2\mu}\over K_{2 \tau_2^{\bot}}}\,
\ve_{2 \tau_2^{\bot}}\,\kappa(K_{2 \tau_2^{\bot}}) +
\left(\ve_{2\mu} - {K_{2\mu}\over K_{2\tau_2^{\bot}}}\, \ve_{2 \tau_2^{\bot}} \right)\,
\kappa(K_{2 \tau_2^{\bot}}/2) \right]
\, e^{-{3\pi\over 8}\,K_{1 \mu}}+ \\
& + &\left. \ve_{2 \tau}\,\kappa(K_{2 \tau})\,e^{-{3\pi\over 8}\,K_{1 \tau}} \right\} \,  ,
\eea
where $K_{2\tau_2^{\bot}} \equiv K_{2e} + K_{2\mu}$ and
$\ve_{2\tau_2^{\bot}} \equiv \ve_{2e} + \ve_{2\mu}$.
This expression for the asymmetry neglects  the flavour coupling effects 
studied in \cite{Antusch:2010ms}.
In the present model only the last term will survive, as we shall justify shortly, and we may 
drastically approximate the above expression to a much simpler one,
\beq\label{twofl2} \nonumber
N_{B-L}^{\rm lep, f} \simeq 
\ve_{2 \tau}\,\kappa(K_{2 \tau})\,e^{-{3\pi\over 8}\,K_{1 \tau}}  \,  ,
\eeq
which says that the only relevant asymmetry is that one produced at the $N_2$ scale in the tauon flavour, 
simply given by $\ve_{2 \tau}\,\kappa(K_{2 \tau})$ with our normalisation,
followed by exponential washout $e^{-{3\pi\over 8}\,K_{1 \tau}}$
the $N_1$ scale.
\footnote{
Of course in models with $M_1 \gtrsim 10^{9}\,{\rm GeV}$  or realising a crossing level solution \cite{afs} with $M_1$ and $M_2$ close enough to have
resonant leptogenesis \cite{resonant}, this 
contribution must also  be considered and the lightest RH neutrino
washout is then not a crucial problem as in the $N_2$-dominated scenario. Examples of realistic $SO(10)$ models 
realising this case were discussed in \cite{mohapatra}.}

The flavoured decay parameters $K_{i\a}$ are defined as
\be\label{Kial}
K_{i\a}\equiv {\G_{i\a}+\overline{\G}_{i\a}\over H(T=M_i)}= 
{|m_{D\a i}|^2 \over M_i \, m_{\star}} \,  ,
\ee
where we introduced the neutrino Dirac mass matrix $m_D \equiv v\, Y^{\nu}_{LR}$. 
The $\Gamma_{i\a}$'s and the $\bar{\Gamma}_{i \a}$'s can be regarded
as the zero temperature limit of the flavoured decay rates into $\a$ leptons, 
$\Gamma (N_i \ra \phi^\dagger \, l_\alpha)$,
and anti-leptons, $\Gamma (N_i \ra \phi \, \bar{l}_\alpha)$, in a three-flavoured regime, where
lepton quantum states can be treated as an incoherent admixture  of the three flavour components.  
The efficiency factors at the production can be calculated using \cite{pedestrians}
\be\label{kappa}
\k(K_{2\a}) = 
{2\over z_B(K_{2\a})\,K_{2\a}}
\left(1-e^{-{K_{2\a}\,z_B(K_{2\a})\over 2}}\right) \,  , \;\; z_B(K_{2\a}) \simeq 
2+4\,K_{2\a}^{0.13}\,e^{-{2.5\over K_{2\a}}} \,   .
\ee
This expression is valid for an initial thermal abundance but, as we will see in a moment, since
the production will prove to occur in the strong wash-out regime, there is actually
independence of the  initial RH neutrino abundance.  Moreover in this case the efficiency factor is well
approximated by $\k(K_{2\a})\simeq 0.5/K_{2\a}^{1.2}$. 

The flavoured $C\! P$ asymmetries,  defined as 
$\ve_{2\a} \equiv - (\overline{\G}_{2\a}-\G_{2\a})/(\G_{2\a}+\overline{\G}_{2\a})$, 
can be calculated in general as \cite{crv}
\bea \label{eq:epsIal}
\varepsilon_{2 \alpha}
        &\,=\,& {3\over 16\,\pi}\,\frac{M_2\,m_{\rm atm}}{v^2}\,\, \,
 \sum\limits_{j \neq 2}  \, \left(
{\cal I}_{2j}^\alpha \, {\xi(M_j^2/M_2^2)}
+ {2\over 3}\,{\cal J}_{2j}^\alpha \, \frac{M_j/M_2}{M_j^2/M_2^2 -1} \right) \, ,
\eea
where we defined \cite{2RHN},
\bea \label{eq:calIJ}
{\cal I}_{2j}^\alpha \equiv
{{\rm Im} \Big[ \big(m_D^\dagger \big)_{i \alpha} \, \big(m_D \big)_{\alpha j}
\big(m_D^\dagger m_D \big)_{ij} \Big]\over M_2\,M_j\,\mtt \, m_{\rm atm}}~,~~
{\cal J}_{2j}^\alpha \equiv {{\rm Im}
\Big[\big (m_D^\dagger \big)_{i \alpha} \, \big(m_D \big)_{\alpha j} \big(m_D^\dagger m_D \big)_{ji} \Big]
\over  M_2\,M_j\,\mtt \, m_{\rm atm}} \, ,
\eea
with $\mtt \equiv (m_D^{\dag}\, m_D)_{2 2}/M_2$, and
\be
\xi(x)=\frac{2}{3}x\left[(1+x)\ln\left(\frac{1+x}{x}\right)-\frac{2-x}{1-x}\right] \,  .
\ee
Terms $\propto {\cal I}_{21}^\alpha\,\xi(M^2_1/M^2_2), {\cal J}_{21}^\alpha,  {\cal J}_{23}^\alpha$
are strongly suppressed in a  way that in the $N_2$-dominated scenario the flavoured $C\!P$ asymmetries 
$\ve_{2\a}$'s  can be approximated by 
\be\label{eps2al}
\ve_{2\a} \simeq  {3\over 16\pi}\frac{\,M_2\,m_{\rm atm}}{v^2}\, {\cal I}_{23}^\alpha  \,  .
\ee
This is because the two terms (for $j=1$) from the interference  with the lightest RH neutrinos 
are $\propto M_1/M_2$,  while the second term from the interference with the heaviest RH neutrinos is 
$\propto M_2/M_3$ and, therefore, they are suppressed compared to ${\cal I}_{23}^\alpha$. 

Let us now calculate ${\cal I}^\alpha_{23}$.  First of all it is immediate to see that
since $Y^{\nu}_{e3}= 0$ then necessarily $\ve_{2e}=0$.  Even though the muon asymmetry does not exactly vanishes
as the electronic one, however, it is suppressed as $\ve_{2\m} \propto a\,b^2\,M'_{13}/M'_3 \sim  10^{-17}$
and can be neglected. 
This implies that, at least at lower order, the observed asymmetry can only be
produced in the tauon flavour, as in Eq.(\ref{twofl2}). 
The expression for the final asymmetry, expressed in terms
of the baryon-to-photon number ratio, then becomes extremely simple,
\be\label{etaBtau}
\eta_B \simeq a_{\rm sph}\,{N_{B-L}^{\rm f}\over N_{\g}^{\rm rec}} \simeq 
0.01\,  \ve_{2\tau} \, \k(K_{2\tau}) \, e^{-{3\pi\over 8}\,K_{1\tau}} \,  ,
\ee
where $a_{\rm sph}=28/79$ is the fraction of $B-L$ asymmetry that is converted 
into a baryon asymmetry by sphaleron processes in equilibrium, $\k(K_{2\t})$ is the 
efficiency factor for the tauon asymmetry at the end of the $N_2$-production and $N_{\g}^{\rm rec}$ is
the number of photons at recombination in the given portion of co-moving volume. 
\footnote{This expression 
is valid independently of the normalisation of the abundances $N_X$, since the normalization 
factor would cancel out in the ratio $N_{B-L}^{\rm f}/N_{\g}^{\rm rec}$.} 

We have first of all to calculate $K_{1\t}$ and check that it is possible to have $K_{1\t} \lesssim 1$.
From the general expression eq.~(\ref{Kial}) one has
\be
K_{1\t} = {|m_{D\t 1}|^2 \over M_1 \, m_{\star}} = {v^2 \, a^2 \over m_{\star}\,M_1}\,
\left| 1 - \g \right|^2  \,  ,
\ee
showing that the quantity $\g$, defined in Eq.~(\ref{gamma}) and originating from the mixing 
parameter $M'_{13}$, plays a crucial role. 
Indeed since $v^2\,a^2/(m_{\star}\,M_1) \sim 20 \gg 1$, the possibility to have $K_{1\t}\lesssim 1$
necessarily relies on having ${\rm Arg}[\g] \simeq 0$ implying $\phi \sim -(\xi + 3\pi/5)$. 

Let us now calculate $\ve_{2\t}$ from the eq.~(\ref{eps2al}). Considering that
$\mtt = 21\,v^2\,b^2/M_2$ and that 
${\cal I}_{23}^\tau \simeq 4\,b^2\,c^2\,\sin(2\,\phi +\xi + 6\,\pi/5) \,v^4/ (M_2\,M_3\,\mtt\,m_{\rm atm})\,$,
one finds
\be\label{eps2taubis}
\ve_{2\tau} \simeq {c^2\over 28\,\pi}\,{M_2 \over M_3}\,\sin(2\phi+\xi + 6\pi/5)  \,  .
\ee
When the condition for $K_{1\t}\lesssim 1$ on $\phi$ is imposed one has
\be
\ve_{2\t} \simeq -{c^2\over 28\,\pi}\,{M_2\over M_3}\,\sin \xi  \,  ,
\label{eps2tau}
\ee
showing that out of the five possible values of $\xi$, only $\xi=-2\pi/5,-4\pi/5$ 
can lead to the correct sign of the asymmetry.  An order-of-magnitude estimation 
gives then $|\ve_{2\t}|\sim 10^{-7}$--$10^{-6}$. 

Finally from the eq.~(\ref{Kial}) we can calculate
\be
K_{2\t} = {|m_{D \t 2}|^2 \over M_2 \, m_{\star}} = {4\over (9)}\,{b^2\,v^2 \over M_2\,m_{\star}} \sim 10\,  ,
\ee 
that, plugged into the eq.~(\ref{etaBtau}) gives $\eta_B \sim 10^{-11}$--$10^{-9}$, showing that
potentially the observed value of the asymmetry could be reproduced. However when one tries to fit
simultaneously the asymmetry and the mixing parameters one finds that the condition $\g \simeq 1$,
necessary to have $K_{1\t} \lesssim 1$, is incompatible with the possibility to reproduce the correct values of 
the mixing parameters so that the asymmetry produced at the $N_2$ scale is afterwards 
completely washed out at the $N_1$ scale. 

\section{Leptogenesis with flavour coupling} 

As we have just seen the asymmetry is mainly  produced by the next-to-lightest RH neutrinos  in the tauon flavour 
but this asymmetry is fully washed-out by the lightest RH neutrinos since the condition $K_{1\t}\lesssim 1$
is not compatible with the measured values of the mixing parameters. 

However, one has also to consider that part of the asymmetry in the tauon flavour is transferred 
to the electron and muon flavours by flavour coupling effects due primarily to the fact that  $N_2$-decays 
produce in addition to an asymmetry in the tauon lepton
doublets also  an  (hyper charge) asymmetry in the Higgs bosons. 
This Higgs asymmetry unavoidably induces, through the inverse decays,  also an asymmetry in the lepton doublets 
that at the production are a coherent admixture of electron and muon components. Therefore, in this case, inverse decays
actually produce an asymmetry instead of wash it out as in a traditional picture. 
A somehow smaller effect is also due to the asymmetries stored into quarks and into right handed charged leptons. 

The account of flavour coupling effects modifies the usually considered expression for the asymmetry
eq.~(\ref{etaBtau}) resulting into \cite{Antusch:2010ms}
\be
\eta_B \simeq \sum_{\a=e,\m,\t} \eta_B^{(\a)} \,  ,
\ee 
where in addition to the tauon contribution eq.~(\ref{etaBtau}) one also has an electron contribution
\be
\eta_B^{(e)} \simeq - 0.01\,  \ve_{2\tau} \, \k(K_{2\tau}) \, {K_{2e}\over K_{2e} + K_{2\mu}}\,
C^{(2)}_{\tau^{\bot}\tau}\,e^{-{3\pi\over 8}\,K_{1e}}   \,   ,
\ee
and a muon contribution given by
\be\label{etaBmu}
\eta_B^{(\mu)} \simeq   - 0.01\,  \ve_{2\tau} \, \k(K_{2\tau}) \left( {K_{2\mu}\over K_{2e} + K_{2\mu}}\,
C^{(2)}_{\tau^{\bot}\tau}\,-{K_{1\m}\over K_{1\t}}\,C^{(3)}_{\mu\tau} \right) \, e^{-{3\pi\over 8}\,K_{1 \mu}} \,  .
\ee
It should be noticed how the source of the electron and muon asymmetries is in any case the tauon asymmetry,
but part of this induces a muon and an electron asymmetry thanks to flavour coupling.  
The flavour coupling coefficients are given by $C^{(2)}_{\tau^{\bot}\tau}= 104/589$ and 
$C^{(3)}_{\mu\tau}=142/537$. Notice that  we are neglecting additional
correcting terms containing products of the flavour coupling coefficients and we are also neglecting terms
$\propto \ve_{2\tau^{\bot}}$ since this is always too small to generate sizeable contributions. 
Also notice that from Eqs.~(\ref{Ynufl}) and (\ref{Kial}) one can see immediately that $K_{1e}=0$.  

Notice moreover that the second term in the muon contribution comes from flavour coupling at the lightest RH neutrino wash-out
that works in the same way as at the production: the lightest RH neutrino wash-out processes acting on muon lepton doublets,
in the presence of a non-vanishing Higgs asymmetry, induce a muon asymmetry. However, this term is basically 
much smaller than the first term that gives the dominant contribution to the total asymmetry, indeed it dominates on the 
electronic term as well since $K_{2e}\ll K_{2\m}$, and is in the end responsible 
for the two solutions that we found. 

We have indeed performed again a numerical fit for all the different cases and this time we have found that, 
both for the case A and for the case B with $\xi=+4\pi/5$, there is indeed a solution, shown in Table 3, with 
an acceptably small  $\chi^2$ value. 
\begin{table}
	\centering
\begin{tabular}{|c| c |c| }
				      \hline
CASE  & A & B \\
\hline \hline
$ \xi $  &   \multicolumn{2}{c|}{+$4\pi/5$}      \\
\hline \hline
$\chi^2_{\rm min}$ & 5.15  & 6.1        	\\
                    \hline
$M'_{11}/{\rm 10^6 GeV}$ & 1.30 &  1.33     \\
			\hline 			
$M'_{22}/{\rm 10^{10} GeV}$  & 0.48 &  4.35      \\
			\hline	
$M'_{33}/{\rm 10^{12} GeV}$ & 2.16 &  1.31     \\
			\hline	
$M'_{13}/{\rm 10^{10} GeV}$  & 1.81 & 0.61      \\
			\hline	
$M'_{13}/M'_{22}$  & 3.75 &  0.141   \\
			\hline	
$\phi/\pi$    & 0.795   &  0.788         \\
			\hline		
$M_{1}/10^7 {\rm GeV}$    &  15  &  2.7      \\
			\hline	
$M_{2}/10^{10} {\rm GeV}$   & 0.483      & 4.35   \\
			\hline	
$M_{3}/10^{12}\, {\rm GeV}$     & 2.16   & 1.31   \\
  			\hline
$|\g|$ &  203 & 38   \\
			\hline	
			\hline
$m_1/{\rm meV}$    & 2.3  & 2.3         \\
			\hline                                        
$m_2/{\rm meV} \;   (p_{\D m^2_{12}})$   &  8.93 (-0.22)  & 8.94  (-0.25)     \\
			\hline 
$m_3/{\rm meV}  \; (p_{\D m^2_{13}})$  & 49.7 (+0.17) &  49.7 (+0.21)      \\
			\hline   
$\sum_i m_i/{\rm meV}$  &  61 &  61       \\
			\hline   			
$m_{ee}/{\rm meV}  $    & 1.95 &  1.95       \\
			\hline  
$\theta_{12}/^{\circ} \; (p_{\theta_{12}})$ & 33.0 (-0.58)   &  33.0 (-0.66)      \\
			\hline   
$\theta_{13}/^{\circ} \; (p_{\theta_{13}})$   & 8.40 (-0.47) &  8.40 (-0.49)     \\
			\hline   
$\theta_{23}/^{\circ} \; (p_{\theta_{23}})$  & 53.3 (+2.1) &  54.0 (+2.3)    \\
			\hline   
$\d/^{\circ}$                   & 20.8 &  23.5   \\
			\hline
$\b_1/^{\circ}$                  &  118 &  115          \\
			\hline
$\b_2/^{\circ}$                  & 281  &  278    \\
			\hline   			 	      
$\a_{21}/^{\circ}$                 &  163  & 162       \\
			\hline   
$\a_{31}/^{\circ}$                 & 242 & 245   \\
			\hline   			 	      
$\rho/^{\circ}$                 &  279  & 279       \\
			\hline   
$\sigma/^{\circ}$                 & 220 & 221   \\

			\hline   \hline	
$ \eta_B /10^{-10} \, (p_{\eta_B})$        & 6.101 (+0.01) &  6.101 (+0.01)    \\
\hline
$ \ve_{2\t} $      &  -$8.1 \times 10^{-6}$  & $ -1.3 \times 10^{-5}$     \\
\hline
$ K_{1\m} $     &  0.11   &  0.58   \\
\hline	
$ K_{1\t} $        & 4341 &  800    \\
\hline	
$ K_{2\t} $         & 7.3 &  7.3    \\
\hline	
$ K_{2\m} $        & 29.2 &  29.2      \\
\hline	
$ K_{2e} $        & 1.8 &  1.8      \\
\hline \hline
$ |(\widetilde{m}_{\nu})_{11}|/{\rm meV} $        & $6.6 \times 10^{-6}$ &  $3.7\times 10^{-5} $     \\ \hline	
$ |(\widetilde{m}^{-1}_{\nu})_{33}|/{\rm meV}^{-1}  $        & 0.22 &   1.2   \\ \hline	
$   |(\widetilde{m}_{\nu})_{12}|/{\rm meV}$        & $2.6 \times 10^{-5}$ &   $1.5 \times 10^{-4}$     \\ \hline	
			
                    \end{tabular} 
			\caption{Solutions found for flavour coupled leptogenesis with $\chi^2_{\rm min}< 100$ (neutrino masses
			are NO).}
		\label{table}
\end{table}
We found no other solution with a $\chi^2$ lower than 100 for all the other cases ($\xi=0, \, \pm 2\pi/5, -4\,\pi/5$) since these
tend to give either a too small $\theta_{12}$ or a too small $\theta_{13}$ or both.
In both cases one can see that $K_{1\m}\lesssim 1$. In this way the flavour coupling induced asymmetry at the production in the muon flavour 
can survive the lightest RH neutrino wash out giving the dominant contribution to the final asymmetry. This is a nice example 
of the ``flavour swap'' scenario envisaged in \cite{Antusch:2010ms}.

It should be noticed that both solutions predict the atmospheric angle well in the second octant, a feature that will be relatively soon tested
by new data from neutrino oscillation experiments. In fact the two solutions give quite similar predictions on $\delta$ and $\theta_{23}$.
In the Figure we have plotted, in the plane $\d$ vs. $\theta_{23}$, points corresponding to solutions with $\chi^2 < 10$ . 
This gives an idea of the allowed region in this plane. It can be seen how for the CASE A the minimum values correspond to 
$(\d,\theta_{23}) \sim (10^{\circ}, 51^{\circ})$ while for the CASE B the whole region is slightly reduced 
(indeed  $\chi^2_{\rm min}\simeq 5$ for the CASE A and $\chi^2_{\rm min}\simeq 6$ for the CASE B)
and shifted to higher values both of $\d$ and $\theta_{23}$ and the minimum
values are $(\d,\theta_{23}) \sim (14^{\circ}, 52^{\circ })$.
This shows that the CASE A is slightly more favoured compared to CASE B. 

It should also be noticed that since now the number of degrees of freedom $\nu =1 $, the $\chi^2_{\rm min}$
can be regarded as an indication of the goodness of fit (g.o.f.), having in mind the previous discussion on the
measurement of $\theta_{23}$. 
Of course we also had the possibility to choose the value of the discrete parameter $\xi$ and between
CASE A and CASE B and this somehow
made things a bit easier, but it is still intriguing that, despite its reduced number of parameters, 
the model can also account  for the matter-antimatter of the Universe.  It should also be said that if the current experimental information on 
$\d$ is taken into account (cf.~eq.(\ref{diracphase})),
 one probably would have an additional contribution $\D \chi^2_\d = p_{\delta}^2 \gtrsim 3$ so that the allowed regions
 shown in the Figure are marginally compatible with current data on $\d$, especially in CASE B while in CASE A the portion
 around the minimum values for $(\d,\theta_{23})$ is still allowed at $\sim 2\s$.  
 \footnote{However, note that in Eq.(\ref{chi2}) we would then have $N=7$ so that the number of degrees of freedom would be 7-5=2 so that one could say that including $\delta$ would actually improve the fit, decreasing 
 $\chi^2/{\rm d.o.f}$, for points with $\D \chi^2 \lesssim 5$. This point is right now quite
indicative since the distribution of $\delta$ is highly non Gaussian.}
\begin{figure}
\vspace*{-0mm}
\hspace*{0mm}
\psfig{file=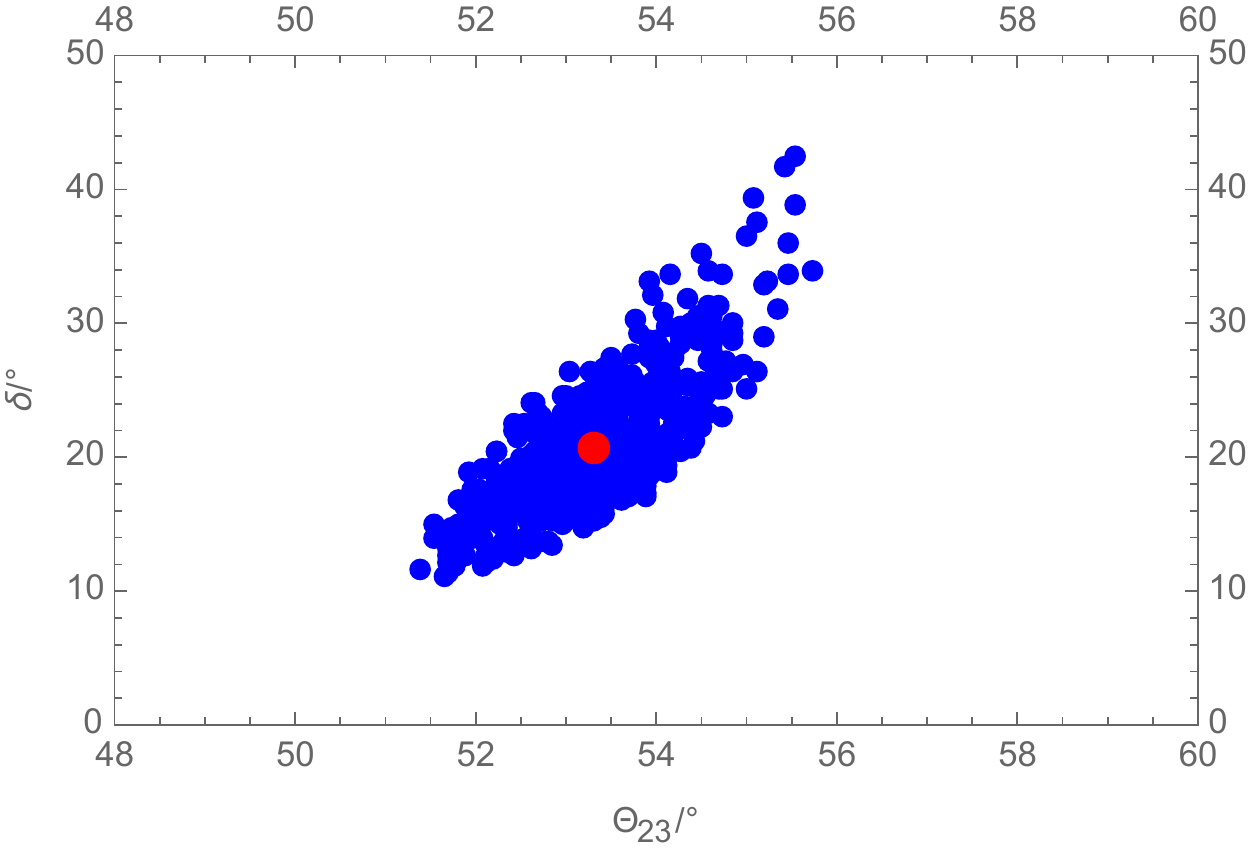,height=60mm,width=75mm}
\hspace{7mm}
\psfig{file=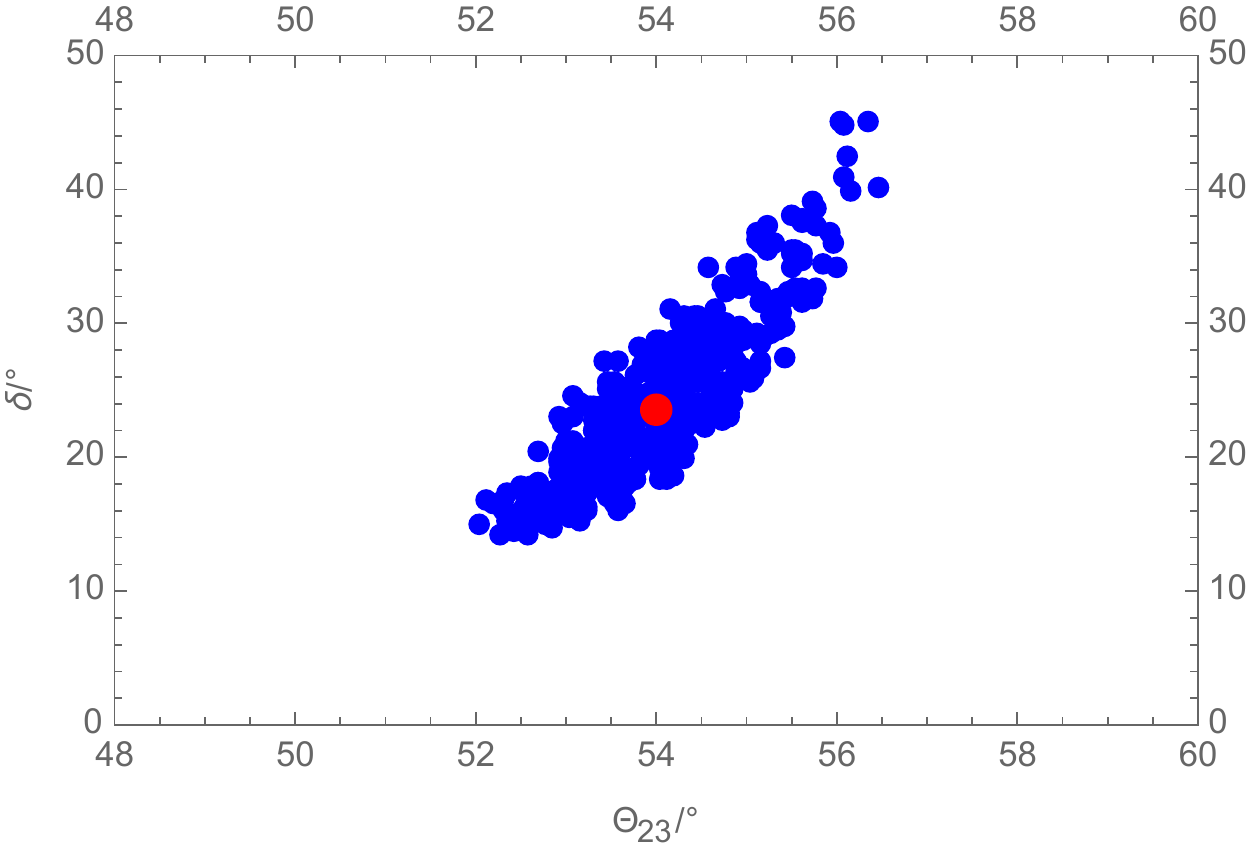,height=60mm,width=75mm}
\caption{Scatter plots of points in the plane $\delta$ vs. $\theta_{23}$ 
with $\chi^2 < 10$ (leptogenesis included) for the CASE A (left) and for the CASE B (right)
for NO and $\xi=+4\pi/5$. The red points correspond to the best fits solutions in Table 3.
These predictions are subject to the theoretical uncertainties discussed in the text.}
\end{figure}
 However, one should also take into account
 that we are currently neglecting corrections from the charge lepton mass matrix that has been approximated to be diagonal and also
 uncertainties on the values of the up quark masses at high scale (see Footnote 1).
 This might help in shifting the allowed regions shown in the Figure toward more favoured experimental values.
  In any case the allowed regions shown in the Figure in the plane $\d$ vs. $\theta_{23}$, in combination with NO, 
  are quite a strong prediction of the model that will be certainly tested during next years.
Much more difficultly testable predictions, certainly not in a close future, are the very small value of the neutrinoless double beta decay effective neutrino mass ($m_{ee} \simeq 2\,{\rm meV}$) and the small increase ($\simeq 3\,{\rm meV}$) of the sum of 
the neutrino masses from its  hierarchial value ($\left.\sum_i m_i\right|_{m_1=0}\simeq 58\,{\rm meV}$).
  
A significant feature of the best fit spectrum for both cases A and B in Table~\ref{table}
is the compressed spectrum of right-handed neutrino masses as compared to the original
estimates in Eq.(\ref{typical}). This stems from the requirement
that successful $N_2$ leptogenesis is proportional to the ratio $M_2/M_3$ in Eq.(\ref{eps2tau})
which must necessarily be larger than naively expected in Eq.(\ref{typical}).
$M_2$ cannot change much,
since according to sequential dominance and the charm quark mass relation
in Eq.(\ref{massrelations}),
it sets the solar neutrino mass scale. It follows that 
the only possibility is for $M_3$ to decrease which requires the parameter $m_c$ to increase
(see Eq.(\ref{massrelations})) and hence the third seesaw matrix 
in Eq.(\ref{seesaw3}), proportional to diag$(0,0,1)$, threatens to dominate the neutrino mass matrix
and pull the atmospheric mixing away from its maximal value.
This threat is averted by noting that, for large $\gamma$, the first matrix in Eq.(\ref{seesaw3})
has a dominant (3,3) element of order $\gamma^2$ which may 
partially cancel the contribution from the third matrix leaving a resulting (3,3) element 
of order $\gamma$, of the same order as the (2,3) and (3,2) elements of the first matrix.
This requires a fine tuning of one part in $\gamma$
(tabulated in Table~\ref{table}).
In order to maintain the correct atmospheric neutrino mass, this increase in $\gamma$ must be compensated by reducing the parameter $m_a$ in proportion to $\gamma$ which, from Eq.(\ref{massrelations}), requires $M_1$ to increase in proportion to $\gamma$. 
In the next section we obtain further
analytic insight into our results from different perspectives.
  
 \section{Analytical insight from general $SO(10)$-inspired leptogenesis}

It is interesting to get some understanding of the numerical results we obtained within the A to Z model from the more general context
of $SO(10)$-inspired leptogenesis \cite{S010inspired,afs,riotto,decription}. In this case  the asymmetry can be calculated analytically assuming that the spectrum of 
the neutrino Dirac mass matrix is hierarchical ($m^D_{\nu 1} \gg m^D_{\nu 2} \gg m^{D}_{\nu 3}$), an assumption certainly
holding in our case.  Here we present a simple generalisation of the results presented in \cite{decription}
providing a very good explanation of the numerical results. 

The starting point is to write the neutrino Dirac mass matrix in the bi-unitary  parameterisation 
(mathematically equivalent to the singular value decomposition of $m^D_{\nu}$),
\be\label{mDbiunitary}
m^D_{\,\nu} = \widetilde{V}_L^{\dagger}\, D_{m^D_{\,\nu}}\,\widetilde{U}_R \,  ,
\ee
where we defined $D_{m^D_{\nu}}\equiv {\rm diag}(m^D_{\nu 1},m^D_{\nu 2},m^D_{\nu 3})$. 
The unitary matrices $\widetilde{V}_L$ and $\widetilde{U}_R$
transform respectively the LH and RH neutrino fields from the flavour basis to the Yukawa basis.  
In our case, from the eq.~(\ref{Ynufl}) for the neutrino Yukawa matrix and parameterising
$\widetilde{V}_L$ analogously to the leptonic mixing matrix as
\be
\widetilde{V}_L = R_{23}(\theta^L_{23})\,R_{13}(\theta^L_{13})\,R_{12}(\theta^L_{12})\,D(\Phi^L) \,   ,
\ee
one has tiny $\theta_{23}^L, \theta^L_{13} \ll 1^{\circ}$ so that
$\widetilde{V}_L \simeq R_{12}(\theta^L_{12})$ with $\theta^L_{12}\simeq 14^{\circ}$. 
Notice that this angle would correspond to the Cabibbo angle $\theta_C$ in the quark sector, that is therefore overestimated
by $\sim 1^{\circ}$. However, turning on down quark mass matrix (non-diagonal) terms, one can reproduce
the correct value.   Similar corrections are expected on the neutrino mixing angles from analogous charged 
lepton mass matrix correcting (non-diagonal) terms. 

Inserting the eq.~(\ref{mDbiunitary}) into the see-saw formula eq.~(\ref{seesaw}) written in the flavour basis, 
one obtains the following expression for the Majorana mass matrix in the Yukawa basis
\be
\widetilde{M}_R \equiv \widetilde{U}^{\star}_R \, D_M \, \widetilde{U}^{\dagger}_R =  
- D_{m^D_{\nu}} \, \widetilde{V}^{\star}_L \, m_{\nu}^{-1} \, \widetilde{V}_L^{\dagger} \, D_{m^D_{\nu}} \,  ,
\ee
and for its inverse
\be\label{Mm1}
\widetilde{M}_R^{-1} \equiv  \widetilde{U}_R \, D_M^{-1} \, \widetilde{U}^{\dagger}_R = 
-D^{-1}_{m^D_{\, \nu}}  \, \widetilde{V}_L \, m_{\nu} \, \widetilde{V}_L^T \, D^{-1}_{m^D_{\, \nu}} \,  .
\ee 
In the above we have transformed the heavy Majorana mass matrix from the diagonal basis 
$M_R=D_M$ in Eq.(\ref{MR}) to the basis 
in Eq.(\ref{mDbiunitary}) in which the Dirac neutrino mass
matrix is diagonal.
The analytical expressions for the RH neutrino masses obtained  in \cite{decription}  in the approximation $\widetilde{V}_L \simeq I$ get extended  for a general $V_L$ simply replacing $m_{\nu} \ra \widetilde{m}_{\nu} \equiv \widetilde{V}_L \, m_{\nu} \, \widetilde{V}_L^T $ \cite{afs}
(the light neutrino mass matrix in the Yukawa basis) obtaining for the RH neutrino masses  
the following analytical expressions
\be\label{Mi}
M_1 \simeq {(m^D_{\,\nu 1})^2 \over |(\widetilde{m}_{\nu})_{11}|} \,    , 
\hspace{5mm}
M_2 \simeq {(m^D_{\,\nu 2})^2 \,|(\widetilde{m}_{\nu})_{11}| \over m_1\,m_2\,m_3\, |(\widetilde{m}^{-1}_{\nu})_{33}|}\,  ,
\hspace{5mm}
M_3 \simeq (m^D_{\,\nu 3})^2 \, |(\widetilde{m}^{-1}_{\nu})_{33}|  \,  .
\ee
In deriving the first and third equalities above we have used the strong up-type quark mass hierarchy
(equal to the eigenvalues of the neutrino Dirac mass matrix in case A), and for the second equality
we have taken the absolute value of the determinant of the seesaw formula.
Plugging the expressions for $M_2$ and $M_3$ into the eq.~(\ref{eps2taubis}) for $\eps_{2\tau}$ 
and in turn this into the eq.~(\ref{etaBmu}) for the dominant muonic contribution to $\eta_B$
and taking into account that $K_{2\m}/(K_{2\m}+K_{2e})=16/17$, one finds 
\be\label{etaBanalbis}
\eta_B \simeq  -  {0.04\,c^2\over 119\,\pi}\,{M_2\over M_3 }\,\k(K_{2\tau}) \,
C^{(2)}_{\tau^{\bot}\tau} \, e^{-{3\pi\over 8}\,K_{1 \mu}}  \,   \sin(2\phi+\xi + 6\pi/5) .
\ee 
From the expressions for $M_2$ and $M_3$ in eq.~(\ref{Mi}) 
and, taking into account the relations~(\ref{Dirac}), (\ref{CASEAbis}) and (\ref{CASEBbis}), 
one can also write for CASE A (CASE B)
\be
\eta_B \simeq -  {0.04 \over 119 \,\pi}\,{(9)\, m^2_{\rm charm}\, |(\widetilde{m}_{\nu})_{11}|\over 
v^2\,m_1\,m_2\,m_3\,|(\widetilde{m}^{-1}_{\nu})_{33}|^2  }\, \k(K_{2\tau}) \,
C^{(2)}_{\tau^{\bot}\tau} \, e^{-{3\pi\over 8}\,K_{1 \mu}}  \,   \sin(2\phi+\xi + 6\pi/5) .
\ee 
Finally one also has
\be\label{K1muanal}
K_{1\mu} = {|m^D_{\nu 21}|^2 \over m_{\star} \, M_1 } \simeq  {(m^{D}_{\nu 2})^2 \over m_{\star}\, M_1} \,
|\widetilde{U}_{R 21}|^2 \simeq {|(\widetilde{m}_{\nu})_{12}|^2\over m_{\star}\,|(\widetilde{m}_{\nu})_{11}|} \,  ,
\ee
where we used the approximations $\sin \theta_C \ll \cos \theta_C \simeq 1$ and 
the following analytical expression for $\widetilde{U}_R$ \cite{afs,decription}
\be\label{URapp}
\widetilde{U}_R \simeq \left( \begin{array}{ccc}
1 & -{m^D_{\nu 1}\over m^D_{\nu 2}} \,  {\widetilde{m}^\star_{\nu 12 }\over \widetilde{m}^\star_{\nu 11}}  & 
{m^D_{\nu 1}\over m^D_{\nu 3}}\,
{ (\widetilde{m}_{\n}^{-1})^{\star}_{13}\over (\widetilde{m}_{\n}^{-1})^{\star}_{33} }   \\
{m^D_{\nu 1}\over m^D_{\nu 2}} \,  {\widetilde{m}_{\nu 12}\over \widetilde{m}_{\nu 11}} & 1 & 
{m^D_{\nu 2}\over m^D_{\nu 3}}\, 
{(\widetilde{m}_{\n}^{-1})_{23}^{\star} \over (\widetilde{m}_{\n}^{-1})_{33}^{\star}}  \\
 {m^D_{\nu 1}\over m^D_{\nu 3}}\,{\widetilde{m}_{\nu 13 }\over \widetilde{m}_{\nu 11}}  & 
- {m^D_{\nu 2}\over m^D_{\nu 3}}\, 
 {(\widetilde{m}_\nu^{-1})_{23}\over (\widetilde{m}_\nu^{-1})_{33}} 
  & 1 
\end{array}\right) \, \begin{pmatrix}
e^{-i\,{\Phi_{1}\over 2}} & 0 & 0 \\
0 & e^{-i\,{\Phi_{2}\over 2}} & 0 \\
0 & 0 & e^{-i\,{\Phi_{3}\over 2}}
\end{pmatrix} ,
\ee
that we generalised here to the case when $\widetilde{V}_L \neq I$ and where the expressions for the phases $\Phi_i$
given in \cite{decription} can be also generalised in terms of $\widetilde{m}_{\nu}$ (we do not need them here).
From these expressions for $\eta_B$ we can now make some considerations
that explain some of the features of the two found numerical solutions 
for $\xi=+4\pi/5$ (see Table 3) providing a useful analytical insight. 

\begin{itemize}
\item Inserting the numerical values for $M_2/M_3$,  $K_{1\mu}$ and $\phi$ given in Table 3 and 4
into the eq.~(\ref{etaBanalbis}), the observed value of $\eta_B$ is indeed reproduced.
\item For one of the possible choices of values for $\xi$, the effective leptogenesis phase is maximal for
$\phi= 3\pi/20 -\xi/2 + n\,\pi$. For the case $\xi = 4\pi/5$ the phase is maximal for $\phi= 3\pi/4 + n\pi$
explaining quite well the best fit values found for $\phi$ (see Table 3).
\item From the eqs.~(\ref{Mi}) we can derive an analytical expression for $M_2/M_3$, the other crucial
parameter determining the value of the asymmetry (cf. eq.~(\ref{etaBanalbis})), finding
\be
{M_2 \over M_3} =  {(m^D_{\,\nu 2})^2 \over (m^D_{\,\nu 3})^2 }\,
{|(\widetilde{m}_{\nu})_{11}| \over m_1\,m_2\,m_3\, |(\widetilde{m}^{-1}_{\nu})_{33}|^2}\,  .
\ee
We have verified that indeed, inserting into this equation the measured values for the mixing angles
and for the solar and atmospheric neutrino mass scales, one obtains the correct value of $M_2/M_3$ 
thanks to a reduction of $|(\widetilde{m}^{-1}_{\nu})_{33}|$ to values $\sim 0.1\,{\rm eV}^{-1}$
from phase cancellations (without cancellations one would have $|(\widetilde{m}^{-1}_{\nu})_{\t\t}|\sim 100\,{\rm eV}^{-1}$) 
and that this is necessarily accompanied by the reduction of $|(\widetilde{m}_{\nu})_{11}|$ 
to values $\sim 10^{-6}\,{\rm eV}$ (without cancellations one would have $|(\widetilde{m}_{\nu})_{11}|\sim 10\,{\rm meV}$)
when a condition $K_{1\mu}\lesssim 24$ is also imposed (it is then much more general than $K_{1\m}\lesssim 1$).  
One can see  that this is indeed what happens 
 from Table 3, where we also show the best fit values for $|(\widetilde{m}^{-1}_{\nu})_{33}|$,  $|(\widetilde{m}_{\nu})_{11}|$ and 
 $|(\widetilde{m}_{\nu})_{12}|$.  It is interesting that this results into  a stable value of the next-to-lightest RH neutrino mass 
($M_2 \simeq 5 \times 10^{9}\,{\rm GeV}$ for the CASE A and  $M_2 \simeq 2 \times 10^{10}\,{\rm GeV}$ for the case B)  the same we obtained in Table 2 without imposing leptogenesis, 
while the heaviest RH neutrino mass  reduces to values  $M_3\simeq 2\times 10^{12}\,{\rm GeV}$.
Correspondingly the lightest RH neutrino mass, though it  does not play a direct role since the asymmetry is $N_2$-dominated, is necessarily forced to grow to values $M_1 \sim 10^{7\div 8}\,{\rm GeV}$.  This is something we have already discussed at the end
of Section 5 starting within the model parameterization and that we have now seen also from a bottom up perspective. 
Moreover the value of the lightest neutrino mass and of the neutrinoless double beta decay effective neutrino 
mass $m_1 \simeq m_{ee} \simeq 2\,{\rm meV}$, in agreement with what we obtained fully numerically in Table 3. 

Therefore, the eqs.~(\ref{Mi}) do explain quite well the obtained results and they also seem to point to a more general result: 
 given the measured values 
of the low energy neutrino parameters, the reduction of $M_3$ generally (the condition 
$K_{1\m}\lesssim 24$ is quite general) implies an uplift of $M_1$ while
$M_2$ remains stable at a scale $\sim 10^{10}\,{\rm GeV}$. In other words the current measurements 
are such that the condition for the realisation of the $M_2-M_3$ crossing level (reduction of $|(\widetilde{m}^{-1}_{\nu})_{33}|$) necessarily implies also the
occurrence of the $M_1-M_2$ crossing level solution, that would lead to a quasi-degenerate RH neutrino mass spectrum
in the close vicinity of the crossing level.
We have checked this statement also verifying that if one let $\theta_{13}$ to be free, then this
effect occurs only for $\theta_{13} \sim 7^{\circ}\div 15 ^{\circ}$. 
In other words, with current values of the neutrino oscillation parameters, a deviation from a strong hierarchy of the RH neutrino masses proportional to the squares of the up-quark masses,
seems to point, for $K_{1\m}\lesssim 24$, toward a RH neutrino spectrum where all three RH neutrino masses tend to get closer.
This analytical insight seems to help understanding also recent numerical results where quark-lepton parameters have been fitted
within $SO(10)$-models either excluding \cite{rodejohann} or including \cite{meloni} leptogenesis and 
either hierarchical or compact RH neutrino mass spectra  have been found but never crossing level solutions with only 
two close RH neutrino masses.  

\item The best fit  values of the flavoured $C\!P$ asymmetry $\ve_{2\t}$ in Table 3, the source
of the asymmetry,  is well above the upper bound \cite{di,upperbound,2RHN}
\be\label{upperbound}
\ve_{1\alpha}^{\cal I} \lesssim {3\over 16\,\pi}\,{M_1\,m_{\rm atm}\over v^2}\,\sqrt{K_{1\alpha}\over K_1} 
\simeq
10^{-6}\,\left({M_1\over 10^{10}\,{\rm GeV}}\right)\,\sqrt{K_{1\alpha}\over K_1}   \,   ,
\ee 
holding for the term in the lightest RH neutrino $C\!P$ flavoured asymmetries analogous to the first term in eq.~(\ref{eq:epsIal}). 
Indeed one has  $|{\cal I}^{\t}_{23}| \simeq 17 \gg 1$ for the best fits of Table 3. This is possible if the absolute values of the 
entries of the orthogonal matrix are much larger than unity. This can be understood considering that
 the orthogonal matrix \cite{casasibarra} in $SO(10)$-inspired models
is given by ($U\equiv U_{PMNS}$)
\be\label{Omegaapp}
\hspace{-15mm}\O \simeq 
\left( \begin{array}{ccc}
- {\sqrt{m_1\,|\widetilde{m}_{\nu 11}|}\over \widetilde{m}_{\nu 11}}\,U_{e 1}  & 
\sqrt{m_2\,m_3\,|(\widetilde{m}_{\nu}^{-1})_{33}| \over |\widetilde{m}_{\nu 11}|}\,
\left(U^{\star}_{\m 1} - U^{\star}_{\t1}\,{{(\widetilde{m}_{\nu}^{-1})_{23}}\over (\widetilde{m}_{\nu}^{-1})_{33}}\right) & 
{U^{\star}_{31}\over \sqrt{m_1\,|(\widetilde{m}_{\nu}^{-1})_{33}|}} \\
- {\sqrt{m_2\,|\widetilde{m}_{\nu 11}|}\over \widetilde{m}_{\nu 11}}\, U_{e 2} & 
\sqrt{m_1\,m_3\,|(\widetilde{m}_{\nu}^{-1})_{33}| \over |\widetilde{m}_{\nu 11}|}\,
\left(U^{\star}_{\m 2} - U^{\star}_{\t2}\,{{(\widetilde{m}_{\nu}^{-1})_{23}}\over (\widetilde{m}_{\nu}^{-1})_{33}}\right)  
& {U^{\star}_{32}\over \sqrt{m_2\,|(\widetilde{m}_{\nu}^{-1})_{33}|}}  \\
- {\sqrt{m_3\,|\widetilde{m}_{\nu 11}|}\over \widetilde{m}_{\nu 11}}\,U_{e 3} & 
\sqrt{m_1\,m_2\,|(\widetilde{m}_{\nu}^{-1})_{33}| \over |\widetilde{m}_{\nu 11}|}\,
\left(U^{\star}_{\m 3} - U^{\star}_{\t 3}\,{{(\widetilde{m}_{\nu}^{-1})_{23}}\over (\widetilde{m}_{\nu}^{-1})_{33}}\right)  
& {U^{\star}_{33}\over \sqrt{m_3\,|(\widetilde{m}_{\nu}^{-1})_{33}|}}  
\end{array}\right)  \,  D_{\Phi},
\ee
where $D_{\Phi}\equiv {\rm diag}(e^{-i{\phi_1\over 2}},e^{-i{\phi_2\over 2}},e^{-i{\phi_3\over 2}})$, an analytical expression that generalises that one given in \cite{decription} when $\widetilde{V}_L\neq I$.
One can see that for the best fit values of  $\widetilde{m}_{11}$ and $\widetilde{m}^{-1}_{33}$ 
one has that all the absolute values of the $\O$ entries in the first and third column are much higher than unity.
This is confirmed by the numerical results we find for $\O$, corresponding to the best fits in Table 3 for CASE A 
\be\label{Omegaapp}
\hspace{-15mm}\O^{({\rm CASE A})} \simeq 
\left( \begin{array}{ccc}
-4.40016-15.9889\, i & 0.0930875 -0.894045\,i  & -16.0396+4.38107\,i \\
-15.9446+3.40333\,i & -1.15394+0.0537137\, i & 3.40494 +15.9553\,i \\
-3.69174+4.35811\, i & 0.709793 +0.204576\,i  &   4.37787 +3.64191\, i  
\end{array}\right)  \,  ,
\ee
and for CASE B
\be\label{Omegaapp}
\hspace{-15mm}\O^{(CASE B)} \simeq 
\left( \begin{array}{ccc}
-1.77835-6.85986\, i & 0.108413 -0.897431\,i  & -6.97828+1.73423 \,i \\
-6.87598+1.34103 \,i & -1.15331+0.0386159\, i & 1.34278 +6.90018\,i \\
-1.64314+1.81259\, i & 0.710523 +0.199612\,i  &   1.85785 +1.52677 \, i  
\end{array}\right)  \,  .
\ee
These specific expressions for $\O$ should give a clear idea of the involved fine tuned cancellations and
explains why one can have $|{\cal I}^{\t}_{23}| \gg 1$ (implying $\ve_{2\t} \gg (3 M_2 m_{\rm atm}/(16\, \pi\,v^2)$)
and more generally why the flavoured $C\!P$ asymmetries
can be enhanced in the vicinity of crossing level solutions \cite{afs} though  RH neutrino masses are still hierarchical
and leptogenesis is far from being  resonant.
However, our novel solution relying on flavour coupling,  represents in this respect quite a large improvement 
compared to the commonly considered solutions based  on $M_1 \gtrsim 10^{9}\,{\rm GeV}$,
since this requires a further  uplift of the lightest RH neutrino mass of at least two orders of magnitude and, therefore,
even higher fine tuned cancellations. 

\item As we have seen the third important ingredient for the existence of the solution is to have $K_{1\m}\lesssim 1$.
 This is crucial in the case A, while in the case B the electron contribution, sixteen times smaller than the muonic one,  
 would be still sufficient (remember that $K_{1e}=0$).
The expression for $K_{1\m}$ eq.~(\ref{K1muanal}) shows that this condition is realised
for $|(\widetilde{m}_{\nu})_{12}|^2 \lesssim m_{\star}\,|(\widetilde{m}_{\nu})_{11}|$ that
is indeed verified for the best fits in Table 3.

\item We can compare our results to those
presented in \cite{riotto} for $\widetilde{V}_L \simeq V_{CKM}$  where flavour coupling was neglected. 
The values of the Majorana phases in Table 3,  match with those in \cite{riotto} though in a marginal region
(for an easier comparison in Table 3 we also give the the Majorana phases in the convention 
${\rm diag}(e^{i\,\rho},1,e^{i\,\s})$ as in \cite{afs,riotto}). 
The same it is true for the values of $\theta_{13},\theta_{23},\d$. In \cite{riotto} it can be also
seen how the ratio $M_2/M_3$ can get reduced around $m_1\simeq 2.5\,{\rm meV}$.  Moreover in \cite{riotto}
there is a lower bound on $M_2 \gtrsim 5 \times 10^{10}\,{\rm GeV}$ that corresponds to 
$\alpha_2 \equiv m^D_{\nu 2}/m_{\rm charm}\gtrsim 3$,
while we find that successful leptogenesis
is obtained for $M_2 \simeq 5 \times 10^{9}\,{\rm GeV}$ and we have $\a_2 =1$ in the CASE A. 
This is because the $C\!P$ asymmetry, thanks to the condition $K_{1\m}$ that makes possible much lower values of $M_2/M_3$,
is greatly enhanced compared to the upper bound eq.~(\ref{upperbound}).  
Moreover  in \cite{riotto} there are no points with $M_1$ uplift. This is explained since there the solutions correspond
to $K_{1\t} \lesssim 1$ while here with flavour coupling the solutions correspond to $K_{1\m}\lesssim 1$ and so there is 
an intrinsic difference.
This shows that the account of flavour coupling indeed opens new solutions enlarging the allowed regions in the
space of parameters, in particular making possible  a reduction in the scale of leptogenesis set by $M_2$. 
\end{itemize}

  \section{Conclusions}

The A to Z model can not only provide a satisfactory fit to all parameters in the leptonic mixing matrix 
but can also reproduce the correct value of the matter-antimatter asymmetry with $N_2$-dominated leptogenesis. 
In this respect it is crucial to  account for flavour coupling effects due to the redistribution of the asymmetry 
in particles that do not participate directly to the
generation of the asymmetry, {\em in primis} the Higgs asymmetry.  
In particular a ``flavour swap'' scenario is realised whereby the asymmetry generated in the tauon flavour 
emerges as a surviving asymmetry dominantly in the muon flavour.
The solution works
even in the simplest case where the neutrino Dirac mass matrix is  equal to the up quark mass matrix.  

Neutrino masses are predicted to be NO, with an atmospheric neutrino mixing angle 
well into the second octant and the Dirac phase $\d\simeq 20^{\circ}$, a set of predictions that will be 
tested in the next years in neutrino oscillation experiments.  We expect these values to be slightly corrected
by charged lepton mass matrix corrections and different theoretical uncertainties (for example 
in the values of the up quark masses at the high scale).
In particular we note that charged lepton mixing corrections,
although small in the A to Z model due to the (1,2) entry of
charged lepton and down quark mass matrices being zero,
may yield atmospheric mixing corrections of order one degree.

In conclusion, the novel solution that we presented involving indispensible flavour coupling,
opens new possibilities for successful leptogenesis within realistic $SO(10)$-inspired models.
Although there is fine tuning given by the parameter $\gamma$,
the level of fine tuning is mild, certainly much smaller than in
traditional quasi-degenerate solutions.
The reason for the fine tuning is that the spectrum of RH neutrinos,
must be compressed as compared to 
the ``natural'' $SO(10)$-inspired spectrum proportional to the squares of up-type quark masses. 
However, whereas  models with $M_1 \gtrsim 10^9\,{\rm GeV}$, commonly considered in the literature, involve 
a very compact mass spectrum, approaching degeneracy, 
the compressed mass spectrum for the $N_2$-dominated case considered here remains very hierarchical.


\label{conclusions}

\section*{Acknowledgements}
PDB and SFK acknowledge partial support 
from the STFC Consolidated ST/J000396/1 and 
the European Union FP7 ITN-INVISIBLES (Marie Curie Actions, PITN- GA-2011-289442).
PDB acknowledges financial support  from the NExT/SEPnet Institute.
We wish to thank for useful discussions K.~Babu, B.~Bajc, M.~Re Fiorentin.

\end{document}